\documentclass[twocolumn,aps,prl,final,superscriptaddress,floatfix]{revtex4}

\usepackage{amssymb}

\usepackage[pdftex]{graphicx}

\newcommand{\blue}{}

\newcommand{\invA}{\mathrm{\AA}^{-1}}
\renewcommand{\deg}{^\circ}

\begin{document}

\title{Giant Kohn anomaly and the phase transition in charge density wave ZrTe$_{3}$}

\author{Moritz Hoesch}

\author{Alexey Bosak}
\affiliation{European Synchrotron Radiation Facility, 6 rue Jules Horowitz, 38043 Grenoble Cedex, France}

\author{Dmitry Chernyshov}
\affiliation{Swiss Norwegian Beamlines at the ESRF, 6 rue Jules Horowitz, 38043 Grenoble Cedex, France}

\author{Helmuth Berger}
\affiliation{Ecole Polytechnique Federale de Lausanne, Institut de physique de la matiere complexe, 1015 Lausanne, Switzerland}

\author{Michael Krisch}
\affiliation{European Synchrotron Radiation Facility, 6 rue Jules Horowitz, 38043 Grenoble Cedex, France}

\date{\today}

\begin{abstract}
A strong Kohn anomaly in ZrTe$_3$ is identified in the mostly transverse acoustic phonon branch along the modulation vector $q_P$ with polarization along the $a^*$ direction. This soft mode freezes to zero frequency at the transition temperature $T_P$ and the temperature dependence of the frequency is strongly affected by fluctuation effects. Diffuse x-ray scattering of the incommensurate superstructure shows a power law scaling of the intensity and the correlation length that is compatible with an order parameter of dimension $n=2$.

\end{abstract}


\maketitle

Dynamical modulations of the spin or charge density are a ubiquitous phenomenon in crystalline metals and form an important ingredient in solid state theory \cite{grunerbook}.
In many materials they remain dynamical at all temperatures and contribute to the rich phenomenology of electronic, magnetic and thermal transport and ordering. Recently, their role has been re-discussed in high $T_c$ super-conducting and colossal magneto-resistance oxides, in connection with the formation of striped charge order {\blue\cite{kivelson96}}. Similar to the stripes, a static charge density wave (CDW) removes spectral weight from the Fermi level, while the superconductivity can be supported by dynamical density waves.  In general, the modulation wave vector $\vec{q}$ can be incommensurate with the underlying lattice and is derived from a nesting geometry of the Fermi surface that provides strong screening at $\vec{q}=2\vec{k}_{F}$. Such a nested Fermi surface is found in particular in low-dimensional crystal structures with chain-like atomic arrangements {\blue\cite{grunerbook}}.

Due to the low-dimensional nature, the phase transition, the so-called Peierls transition, to the statically modulated ground state of the one-dimensional chain is strongly affected by fluctuations \cite{lee73}. 
Inelastic neutron scattering studies revealed a giant Kohn anomaly in KCP \cite{renker73} and blue bronze \cite{pouget91}, whereas (TaSe$_4$)$_2$I  \cite{fujishita86,lorenzo98} displayed a temperature dependent soft mode, which did, however, not attain zero frequency at the Peierls transition temperature $T_{P}$. An inelastic x-ray scattering study of the prototypical 1D compound NbSe$_3$ failed to show any clear Kohn anomaly but only a broadening of an acoustic phonon mode at the {\blue first} modulation vector $\vec{q}_1$ {\blue of NbSe$_3$}  \cite{requardt02}.

The modulated ground state itself is readily observed by the appearance of new diffraction peaks, with a position in the reciprocal lattice that represents the modulation vector $\vec{q}$. Similar to other structural phase transitions, the temperature dependence of these superstructure reflections is used to study the ordering of the CDW modulation. 
Again, only a small number of 1D inorganic chain systems show this transition and very few studies exist.
These find that the Peierls transition is governed by a critical exponent of the order parameter, which is compatible with an order parameter of dimension $n=2$ (blue bronze \cite{girault89}) or close to being a (blurred) first order transition ((TaSe$_4$)$_2$I \cite{requardt96}). A clear assignment to a particular universality class has, however, not been found and the question remains open whether the transition can be classified in this way.

While the CDW ground state and its many fascinating properties are well described by theory \cite{grunerbook}, the phase transition is accessible only by experiment. 
In this letter we present an experimental study of the fluctuations and the soft mode phonon in ZrTe$_3$ to elucidate the properties of the charge density wave in the important regime close to and above the phase transition, where the density waves are of dynamical nature.

ZrTe$_3$ crystallizes in a monoclinic structure with space-group $P2_1/m$ (lattice parameters $a=5.89$~\AA, $b=3.93$~\AA, $c=10.09$~\AA, $\alpha=\gamma=90\deg$, $\beta = 97.8\deg$). (ZrTe$_3$)$_{\infty}$ chains similar to NbSe$_3$ run along $b$. Only one type of chain is present in the unit cell. The Peierls modulation $\vec{q}_{P} = (0.07,\ 0,\ 0.333)$ (reciprocal lattice units) \cite{eaglesham84} is transverse to these prismatic chains with a small $a^*$ component and a tripling of the unit cell along the layering $c^*$ direction. The modulation is thus very different from NbSe$_3$, and mostly in the almost equidistant Te-Te chains along $a$, where the prismatic chains are laterally bonded. Resisitvity measurements show an anomaly due to the transition at $T_P=63$~K and the system remains metallic below $T_P$ \cite{takahashi83}. Electronic structure calculations \cite{felser98, stowe98} and a detailed photo-emission study \cite{yokoya05} explain this behaviour by nesting in a small electron pocket of the highly directional Te 5$p_x$ band running along these Te-Te chains, while other sheets of the Fermi surface remain unaffected by the transition. The opening of a gap in the electronic dispersion of this band was found to follow a BCS model of the Peierls transition with a mean field transition temperature $T_{MF}$ about four times higher than $T_P$ \cite{yokoya05}.  Last, but not least, the phase transition gives rise to a very small specific heat anomaly at $T_P$ \cite{chung93}.

The experiment was performed at the inelastic x-ray scattering (IXS) beamline ID28 at the European Synchrotron Radiation Facility. Phonon dispersions were measured with an energy resolution $\delta E= 1.7$~meV. The momentum resolution was $\delta Q=0.05\ \invA$ in the horizontal $a^*-b^*$ plane and $0.15\ \invA$ in the vertical $c^*$ direction. Diffuse scattering and diffraction scans were recorded with an energy-integrating detector at the same wavelength of $\lambda = 0.57$~\AA{} and similar momentum resolution. The samples were grown by chemical vapour transport and formed platelets of $1\times2\times 0.1$~mm$^3$ {\blue along $a$, $b$ and $c$ respectively}, that were glued to the copper cold finger of a closed cycle cryostat using silver paste. The absolute temperature scale was calibrated relative to a diode thermometer mounted on the cold finger by setting the inflection point of the intensity of critical 
scattering at $T_P$ to the known $T_P = 63$~K (see Fig.~\ref{Fig4}).

\begin{figure}
\vspace{6mm}
\centerline{\includegraphics[width = .33\textwidth]{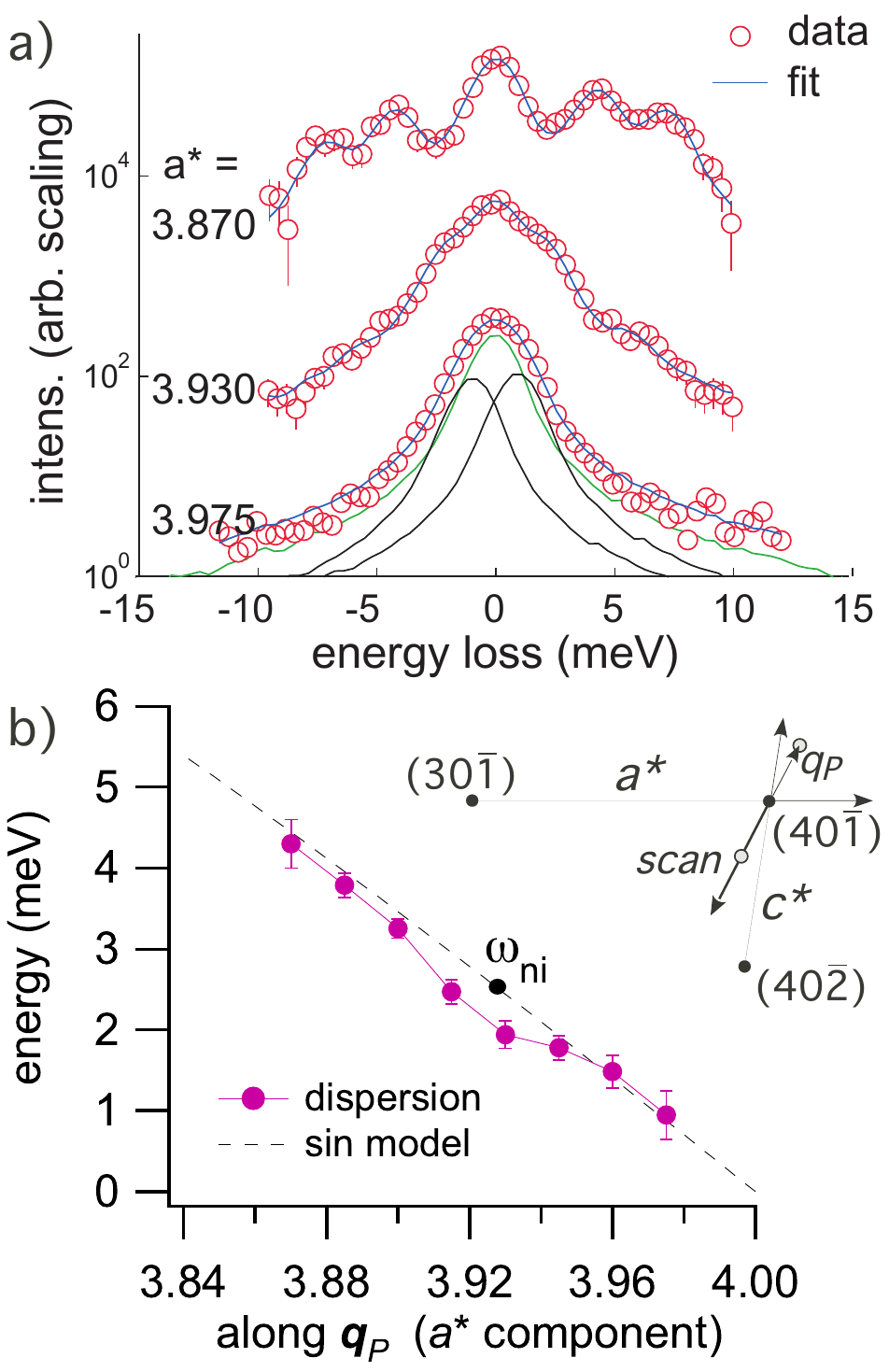}}
\caption{Acoustic phonon along $\vec{q}_{P}$ at $T=100$~K. (a) Representative IXS spectra at momenta close to $(40\overline{1})$ at $\vec{q}_{P}$ ($a^*=3.93$) and across $\vec{q}_{P}$ on a logarithmic scale with arbitrary offset. Error bars are smaller than the symbols except for the outer data points. Solid lines below that data are (shifted) resolution functions that were summed to fit the data. (b) Phonon dispersion extracted from fits to the IXS data. The dashed line is a sinusoidal model dispersion $\omega_{ni}(\vec{q})$ that serves as a guide to the eye. {\blue The inset shows a portion of reciprocal space with the scan direction through $\vec{q}_{P}$ indicated.}}
\label{Fig1}
\end{figure}

The phonon dispersion along the CDW modulation vector was determined as shown in Fig.~\ref{Fig1}. 
IXS spectra were recorded for each momentum transfer $\vec{Q}$ close to the strong Bragg spot $\vec{G}=(40\overline{1})$, {\blue along a straight line from $(40\overline{1})$ through $\vec{q}_{P}$ and beyond, as shown in the inset of Fig.~\ref{Fig1}. In the figure, the $a^*$ component of $\vec{Q}$ is indicated in relative lattice units.} At a temperature $T=100\ \mathrm{K}=1.6 \cdot T_P$, the spectra are dominated by a central peak at zero energy loss for all momenta, in particular close to $\vec{q}_{P}$ ($a^*=3.93$) and on approaching the $\Gamma$-point (lowest curve). 
The spectra are fit by a sum of three or five terms for one or two phonons, respectively. The first term is the central peak, the following terms are phonons in Stokes- and anti-Stokes positions with their intensity ratio given by the detailed balance at the corresponding temperature. No convolution was performed, and the model function was a sum of shifted resolution functions according to these terms, thereby assuming that broadening effects will be small with respect to the experimental resolution, as they cannot be resolved in the strongly overlapping peaks. This method yielded smooth and self-consistent phonon dispersions.

\begin{figure}[tb!]
\centerline{\includegraphics[width = .42\textwidth]{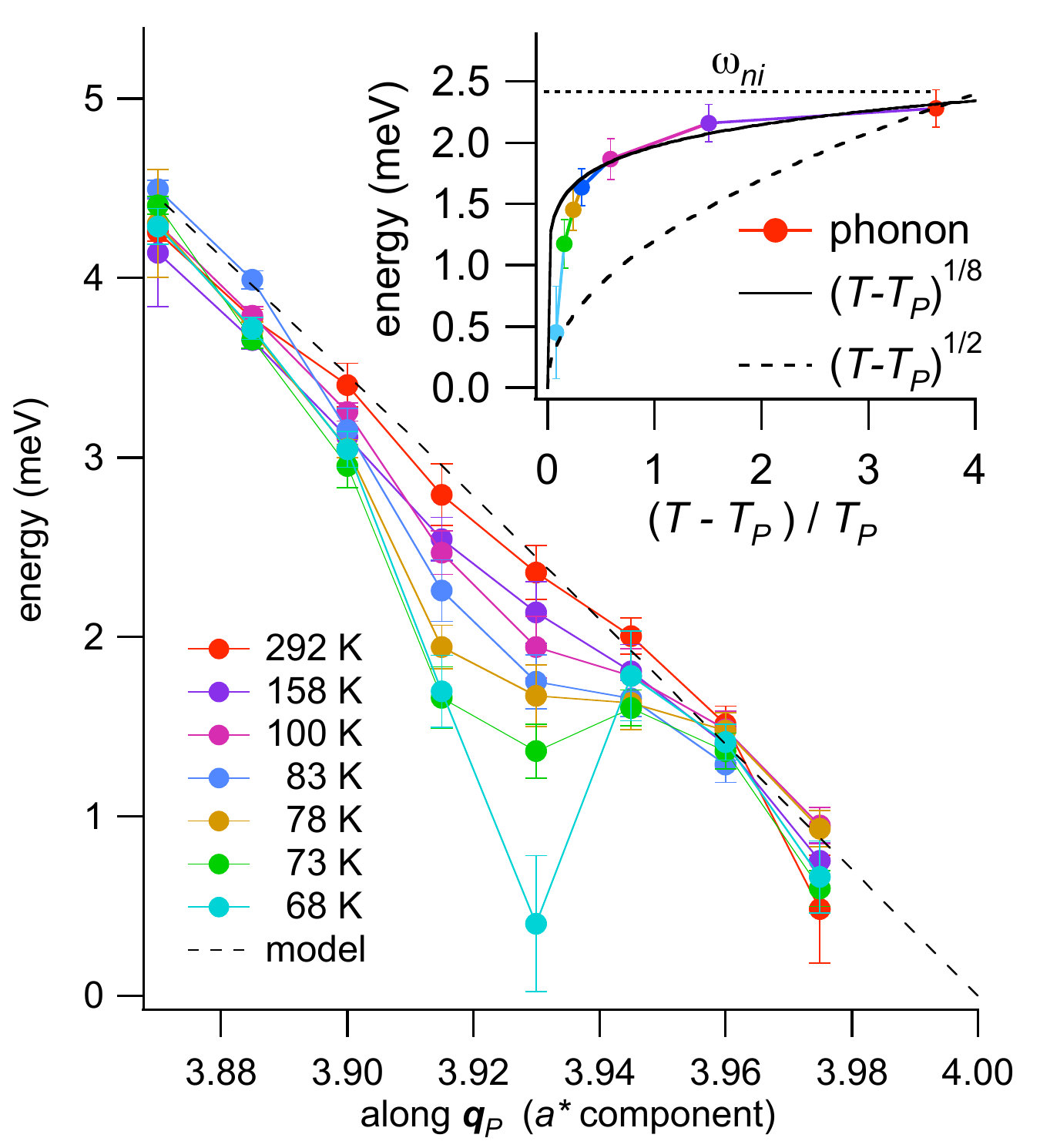}}
\caption{Phonon dispersion from $(40\overline{1})$ in the direction of $\vec{q}_{P}$ at various temperatures $T>T_{P}$ above the transition. The inset shows the phonon energy $\hbar \omega_q$ at $\vec{q}_{P}$ as a function of reduced temperature $(T-T_P)/T_P$. The solid and  dashed lines represent a 1/8 and a 1/2 power law, respectively. }
\label{Fig2}
\end{figure}

\begin{figure*}[t!]
\centerline{\includegraphics[width = 0.99\textwidth]{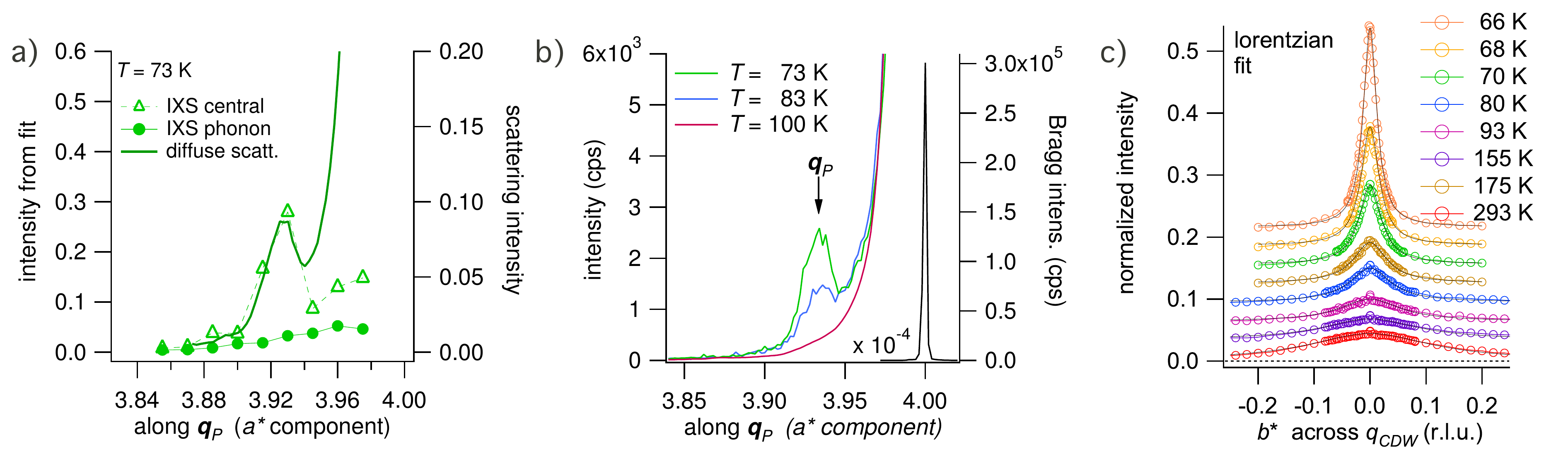}}
\caption{Diffuse scattering along $\vec{q}_{P}$ (a) Symbols represent the intensities extracted from data fitting of the IXS spectra: central peak (triangles), phonon contribution (filled circles). The solid line is the total diffuse scattering intensity as measured with an energy integrating detector.  (b) Diffuse scattering intensities at three selected temperatures as indicated. (c) Scans of diffuse scattering across $\vec{q}_{P}$ along $b^*$ at various temperatures above $T_P$.}
\label{Fig3}
\end{figure*}

The thus extracted dispersion is displayed in Fig.~\ref{Fig1}b). It shows a dip  at $\vec{q}_{P}$ deviating from a sinusoidal model dispersion. This Kohn anomaly (KA) is a direct manifestation of the Fermi surface nesting vector $2k_F$. No such anomaly was observed for other acoustic phonons; in particular a scan along the high symmetry direction $a^*$ (longitudinal acoustic phonon) did not show any anomaly, although the Fermi surface could also afford a certain nesting along this direction. All other modes that were studied did not show any anomalies, and also a Raman scattering study of the phonon modes at the $\Gamma$-point did not show any anomalous temperature dependence \cite{zwick80}. The line along $\vec{q}_{P}$ corresponds to an almost transverse dispersion along the large $c^*$ component. The phonon polarization is along $a^*$. Thus, the displacement pattern of the soft-mode is of mostly transverse character with respect to the Te-Te chains whose Te $5p_x$ orbitals form the quasi 1-dimensional Fermi surface.

The temperature dependence of the Kohn anomaly is traced in Fig.~\ref{Fig2}. As mentioned above, already at room temperature ($4.6 \cdot T_P$) a slight dip is visible, barely outside the experimental error bars. As the temperature is lowered, the dip gets more pronounced, until the phonon frequency drops rapidly to zero at $T_P$. Mean field theory predicts this freezing of the phonon as a logarithmic divergence that would correspond to a square-root power law close to the phase transition \cite{grunerbook}. {\blue Inspection of our data by plotting on a double-logarithmic scale did not reveal a well-defined power-law behaviour. Over a broad range in temperature, the closest correspondence to a power law approximately $(T-T_P)^{1/8}$. On the linear scale in Fig.~\ref{Fig2}, this 1/8 power law and the theoretical expectation of a 1/2 power law are shown.}
Thus, we conclude that the soft mode freezes in the fashion of a giant Kohn anomaly. {\blue Our results suggest that the phonon frequency drops to zero at the phase transition. Above the transition, fluctuation effects lead to a much more rapid recovery of the non-interacting phonon frequency than the logarithmic dependence predicted by mean field theory.}

The deviation from mean field behaviour led us to investigate the thermodynamic  signature of the phase transition by the study of the diffuse scattering ($T>T_P$) and superstructure intensity ($T<T_P$) at $\vec{q}_{P}$. Again, the experimental difficulty lies in the close proximity to the Bragg reflection due to the small component $a^*$ of $\vec{q}_{P}$. A survey of momentum space slightly above $T_P$ showed that the diffuse scattering signal is present for $\vec{Q}=\vec{G}\pm\vec{q}_{P}$ when $\vec{G}$ has a strong component along the $a^*$ direction of the reciprocal lattice. 

Our combined diffuse and inelastic x-ray scattering study allows disentangling the static and dynamical contributions to the diffuse intensity.
On inspection the IXS spectra show that the dominant contribution is  the central peak (zero energy loss). Figure~\ref{Fig3}a) compares the intensity of the central peak and the phonon (symbols) with the energy integrated signal (line) at $T=73\ {\mathrm K}\ >T_P$. The central peak is not only dominant, but its momentum dependence corresponds to that of the diffuse scattering signal. Thus we can interpret the diffuse scattering as a quasi-static diffraction feature which is representative of the (fluctuating) order of the CDW.

\begin{figure}[b!]
\centerline{\includegraphics[width = .5\textwidth]{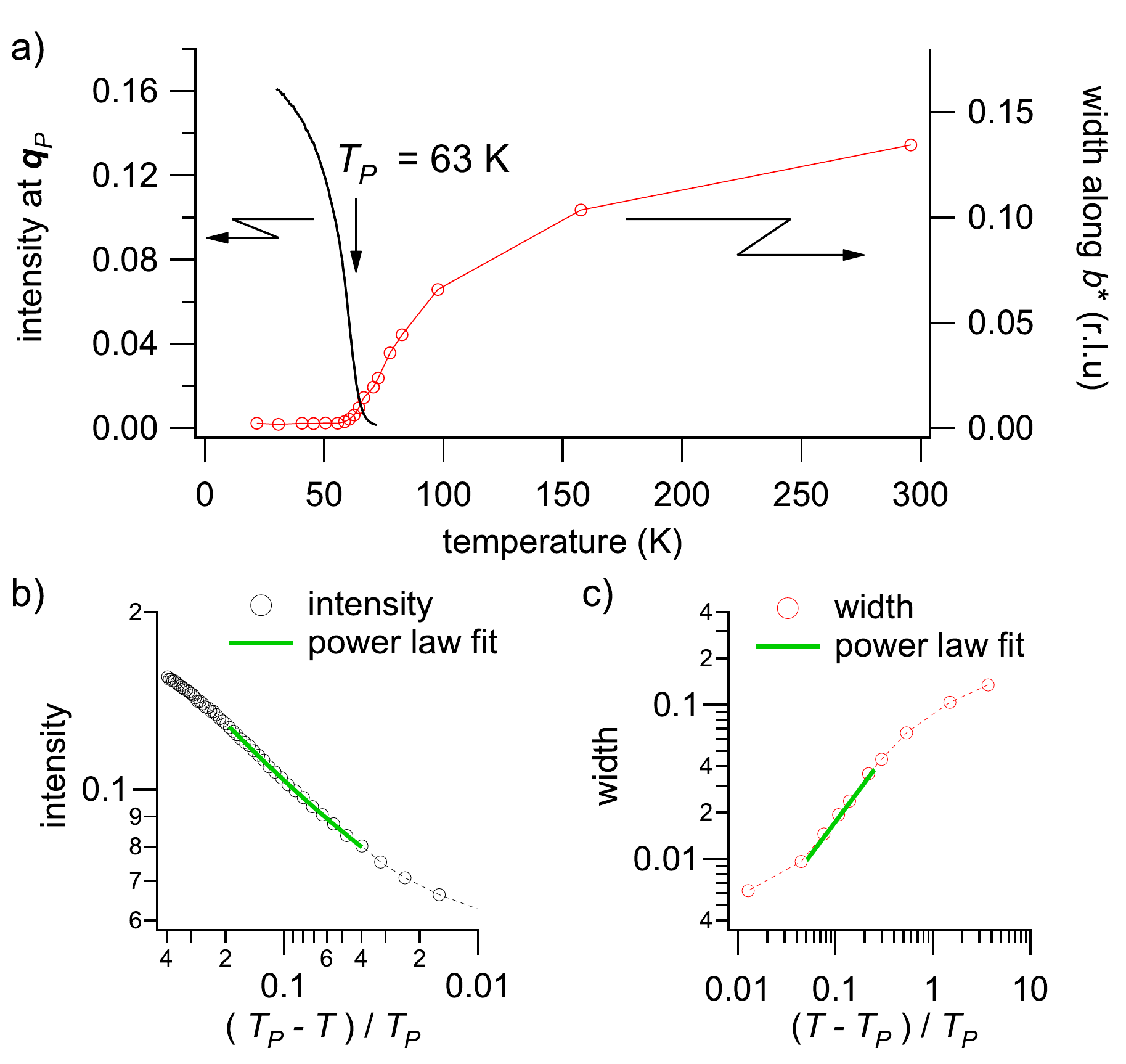}}
\caption{(a) Temperature dependence of superstructure intensity at $-$ (left scale) and the width of the diffuse scattering along $b*$ (right scale). (b) and (c) Double logarithmic representations of the intensity and the width, respectively, versus the reduced temperature. } 
\label{Fig4}
\end{figure}

Scans of the integrated scattering intensity along $\vec{q}_{P}$ are displayed in fig.~\ref{Fig3}b) and show the emergence of a diffuse scattering feature. Due to the strong background of thermal diffuse scattering (TDS) from the acoustic phonon (dominating at high temperatures), the peak shape is not easily determined. A reliable measurement, however, is possible for scans through $\vec{q}_{P}$ along $b^*$ as the rocking curve along this direction is sharp and no resolution broadening affects the width of the intensity profiles. Again, at high temperature a broad peak is partially due to TDS, but the emergence of a sharp feature with decreasing width at lower $T$ is due to the longer correlation length of the CDW order. Below $T_P$ this peak does not sharpen any further, although it remains slightly broadened with respect to the experimental resolution. Its intensity grows rapidly on lowering the temperature below $T_P$.

The temperature evolution of the intensity  and the width along $b^*$ are shown in Fig.~\ref{Fig4}. 
The transition appears slightly broadened around $T_P$. Outside of this range, the intensity and width show power law scaling with reduced temperature. The critical exponents derived from these are $2\beta = 0.26\pm0.06$ for the intensity, that is taken here as a measure of the order parameter squared, and $\nu =0.85\pm 0.2$ for the width, that is proportional to the inverse correlation length. The diffuse scattering peaks as well as the diffraction profiles below $T_P$ were always well represented by a single Lorentzian function. Other critical exponents could not be derived, in particular the shape of the diffuse scattering peak along $a^*$ and $c^*$ is not accessible due to its large width on a sloping strong background. 

The critical exponent $\beta = 0.13 \pm 0.03$ is even smaller than the ones observed in other CDW materials ($0.32\pm 0.05$ for blue bronze \cite{girault89} and $0.23 \pm 0.02$ for (TaSe$_4$)$_2$I  \cite{requardt96}). Thus, a mean field description of the phase transition is not appropriate and fluctuation corrections are needed. Such small exponents $\beta$ of the order parameter are best compatible with a two-dimensional order parameter. The diffuse scattering spot is well defined in momentum space and the Kohn Anomaly is sharp in a three dimensions, thus above $T_P$ a three-dimensional fluctuation regime is observed. Furthermore the exponent $\nu$ of the correlation length is compatible with a three-dimensional XY universality class~\cite{baker78}. 
On the other hand, the Peierls transition could be considered as a discontinuous freezing (1st order) that is blurred by the presence of defects and thus rendered continuous. This would also lead to a power law with a small exponent~\cite{requardt96}. However, our experiment did not show any hysteresis in temperature scans and only a slight specific heat anomaly was reported in ZrTe$_3$~\cite{chung93}. Thus, we conjecture that the phase transition is continuous (2nd order) or very close to that.

In summary, we have identified the soft phonon mode that stabilizes the charge density modulation in ZrTe$_3$ and measured its temperature dependence. It is an acoustic $a^*$ polarized phonon branch with a mostly transverse dispersion along $c^*$. The Kohn anomaly is observed already at high temperatures of 3-4 $T_P$ and only for a sharp momentum space region around $\vec{q}_{P}$. On approaching the phase transition it becomes giant and the phonon freezes into a static modulation of the lattice. The temperature dependence of the softening does not follow mean field predictions. The analysis of superstructure diffraction intensity and diffuse scattering correlation lengths lead to the conclusion that the phase transition is continuous, but governed by an order parameter of dimension $n=2$.

We wish to thank D. Gambetti and P. Dideron for technical assistance. Fruitful discussion with R. Currat, P. Monceau, S.V. Maleyev and J.-P. Pouget are gratefully acknowledged. We also thank Swiss-Norwegian beam lines at ESRF for the beam-time allocation necessary for preliminary mapping of reciprocal space. This work was performed at the European Synchrotron Radiation Facility.

\bibliography{kohn}

\begin{thebibliography}{18}
\expandafter\ifx\csname natexlab\endcsname\relax\def\natexlab#1{#1}\fi
\expandafter\ifx\csname bibnamefont\endcsname\relax
  \def\bibnamefont#1{#1}\fi
\expandafter\ifx\csname bibfnamefont\endcsname\relax
  \def\bibfnamefont#1{#1}\fi
\expandafter\ifx\csname citenamefont\endcsname\relax
  \def\citenamefont#1{#1}\fi
\expandafter\ifx\csname url\endcsname\relax
  \def\url#1{\texttt{#1}}\fi
\expandafter\ifx\csname urlprefix\endcsname\relax\def\urlprefix{URL }\fi
\providecommand{\bibinfo}[2]{#2}
\providecommand{\eprint}[2][]{\url{#2}}

\bibitem[{\citenamefont{Gr\"uner}(1994)}]{grunerbook}
\bibinfo{author}{\bibfnamefont{G.}~\bibnamefont{Gr\"uner}},
  \emph{\bibinfo{title}{Density Waves in Solids}}, vol.~\bibinfo{volume}{89} of
  \emph{\bibinfo{series}{Frontiers in Physics}} (\bibinfo{publisher}{Perseus
  publishing, Cambridge MA}, \bibinfo{year}{1994}).

\bibitem[{\citenamefont{Kivelson and Emery}(1996)}]{kivelson96}
\bibinfo{author}{\bibfnamefont{S.}~\bibnamefont{Kivelson}} \bibnamefont{and}
  \bibinfo{author}{\bibfnamefont{V.}~\bibnamefont{Emery}},
  \bibinfo{journal}{Synth. Metals} \textbf{\bibinfo{volume}{80}},
  \bibinfo{pages}{151} (\bibinfo{year}{1996}).

\bibitem[{\citenamefont{Lee et~al.}(1973)\citenamefont{Lee, Rice, and
  Anderson}}]{lee73}
\bibinfo{author}{\bibfnamefont{P.}~\bibnamefont{Lee}},
  \bibinfo{author}{\bibfnamefont{T.}~\bibnamefont{Rice}}, \bibnamefont{and}
  \bibinfo{author}{\bibfnamefont{P.}~\bibnamefont{Anderson}},
  \bibinfo{journal}{Phys. Rev. Lett} \textbf{\bibinfo{volume}{31}},
  \bibinfo{pages}{462} (\bibinfo{year}{1973}).

\bibitem[{\citenamefont{Renker et~al.}(1973)\citenamefont{Renker, Rietschel,
  Pintschovius, Gl\"aser, Br\"uesch, Kuse, and Rice}}]{renker73}
\bibinfo{author}{\bibfnamefont{B.}~\bibnamefont{Renker}},
  \bibinfo{author}{\bibfnamefont{H.}~\bibnamefont{Rietschel}},
  \bibinfo{author}{\bibfnamefont{L.}~\bibnamefont{Pintschovius}},
  \bibinfo{author}{\bibfnamefont{W.}~\bibnamefont{Gl\"aser}},
  \bibinfo{author}{\bibfnamefont{P.}~\bibnamefont{Br\"uesch}},
  \bibinfo{author}{\bibfnamefont{D.}~\bibnamefont{Kuse}}, \bibnamefont{and}
  \bibinfo{author}{\bibfnamefont{M.~J.} \bibnamefont{Rice}},
  \bibinfo{journal}{Phys. Rev. Lett.} \textbf{\bibinfo{volume}{30}},
  \bibinfo{pages}{1144} (\bibinfo{year}{1973}).

\bibitem[{\citenamefont{Pouget et~al.}(1991)\citenamefont{Pouget, Hennion,
  Escribe-Filippini, and Sato}}]{pouget91}
\bibinfo{author}{\bibfnamefont{J.~P.} \bibnamefont{Pouget}},
  \bibinfo{author}{\bibfnamefont{B.}~\bibnamefont{Hennion}},
  \bibinfo{author}{\bibfnamefont{C.}~\bibnamefont{Escribe-Filippini}},
  \bibnamefont{and} \bibinfo{author}{\bibfnamefont{M.}~\bibnamefont{Sato}},
  \bibinfo{journal}{Phys. Rev. B} \textbf{\bibinfo{volume}{43}},
  \bibinfo{pages}{8421} (\bibinfo{year}{1991}).

\bibitem[{\citenamefont{Fujishita et~al.}(1986)\citenamefont{Fujishita,
  Shapiro, Sato, and Hoshino}}]{fujishita86}
\bibinfo{author}{\bibfnamefont{H.}~\bibnamefont{Fujishita}},
  \bibinfo{author}{\bibfnamefont{S.~M.} \bibnamefont{Shapiro}},
  \bibinfo{author}{\bibfnamefont{M.}~\bibnamefont{Sato}}, \bibnamefont{and}
  \bibinfo{author}{\bibfnamefont{S.}~\bibnamefont{Hoshino}},
  \bibinfo{journal}{Journal of Physics C: Solid State Physics}
  \textbf{\bibinfo{volume}{19}}, \bibinfo{pages}{3049} (\bibinfo{year}{1986}).

\bibitem[{\citenamefont{Lorenzo et~al.}(1998)\citenamefont{Lorenzo, Currat,
  Monceau, Hennion, Berger, and Levy}}]{lorenzo98}
\bibinfo{author}{\bibfnamefont{J.~E.} \bibnamefont{Lorenzo}},
  \bibinfo{author}{\bibfnamefont{R.}~\bibnamefont{Currat}},
  \bibinfo{author}{\bibfnamefont{P.}~\bibnamefont{Monceau}},
  \bibinfo{author}{\bibfnamefont{B.}~\bibnamefont{Hennion}},
  \bibinfo{author}{\bibfnamefont{H.}~\bibnamefont{Berger}}, \bibnamefont{and}
  \bibinfo{author}{\bibfnamefont{F.}~\bibnamefont{Levy}},
  \bibinfo{journal}{Journal of Physics: Condensed Matter}
  \textbf{\bibinfo{volume}{10}}, \bibinfo{pages}{5039} (\bibinfo{year}{1998}).

\bibitem[{\citenamefont{Requardt et~al.}(2002)\citenamefont{Requardt, Lorenzo,
  Monceau, Currat, and Krisch}}]{requardt02}
\bibinfo{author}{\bibfnamefont{H.}~\bibnamefont{Requardt}},
  \bibinfo{author}{\bibfnamefont{J.~E.} \bibnamefont{Lorenzo}},
  \bibinfo{author}{\bibfnamefont{P.}~\bibnamefont{Monceau}},
  \bibinfo{author}{\bibfnamefont{R.}~\bibnamefont{Currat}}, \bibnamefont{and}
  \bibinfo{author}{\bibfnamefont{M.}~\bibnamefont{Krisch}},
  \bibinfo{journal}{Phys. Rev. B} \textbf{\bibinfo{volume}{66}},
  \bibinfo{pages}{214303} (\bibinfo{year}{2002}).

\bibitem[{\citenamefont{Girault et~al.}(1989)\citenamefont{Girault, Moudden,
  and Pouget}}]{girault89}
\bibinfo{author}{\bibfnamefont{S.}~\bibnamefont{Girault}},
  \bibinfo{author}{\bibfnamefont{A.~H.} \bibnamefont{Moudden}},
  \bibnamefont{and} \bibinfo{author}{\bibfnamefont{J.~P.}
  \bibnamefont{Pouget}}, \bibinfo{journal}{Phys. Rev. B}
  \textbf{\bibinfo{volume}{39}}, \bibinfo{pages}{4430} (\bibinfo{year}{1989}).

\bibitem[{\citenamefont{Requardt et~al.}(1996)\citenamefont{Requardt, Kalning,
  Burandt, Press, and Currat}}]{requardt96}
\bibinfo{author}{\bibfnamefont{H.}~\bibnamefont{Requardt}},
  \bibinfo{author}{\bibfnamefont{M.}~\bibnamefont{Kalning}},
  \bibinfo{author}{\bibfnamefont{B.}~\bibnamefont{Burandt}},
  \bibinfo{author}{\bibfnamefont{W.}~\bibnamefont{Press}}, \bibnamefont{and}
  \bibinfo{author}{\bibfnamefont{R.}~\bibnamefont{Currat}},
  \bibinfo{journal}{Journal of Physics: Condensed Matter}
  \textbf{\bibinfo{volume}{8}}, \bibinfo{pages}{2327} (\bibinfo{year}{1996}).

\bibitem[{\citenamefont{Eaglesham et~al.}(1984)\citenamefont{Eaglesham, Steeds,
  and Wilson}}]{eaglesham84}
\bibinfo{author}{\bibfnamefont{D.~J.} \bibnamefont{Eaglesham}},
  \bibinfo{author}{\bibfnamefont{J.~W.} \bibnamefont{Steeds}},
  \bibnamefont{and} \bibinfo{author}{\bibfnamefont{J.~A.}
  \bibnamefont{Wilson}}, \bibinfo{journal}{J. Phys. C}
  \textbf{\bibinfo{volume}{17}}, \bibinfo{pages}{L697} (\bibinfo{year}{1984}).

\bibitem[{\citenamefont{Takahashi et~al.}(1983)\citenamefont{Takahashi,
  Sambongi, and Okada}}]{takahashi83}
\bibinfo{author}{\bibfnamefont{S.}~\bibnamefont{Takahashi}},
  \bibinfo{author}{\bibfnamefont{S.}~\bibnamefont{Sambongi}}, \bibnamefont{and}
  \bibinfo{author}{\bibfnamefont{S.}~\bibnamefont{Okada}}, \bibinfo{journal}{J.
  Phys. (Paris) Colloq.} \textbf{\bibinfo{volume}{44}}, \bibinfo{pages}{C3}
  (\bibinfo{year}{1983}).

\bibitem[{\citenamefont{Felser et~al.}(1998)\citenamefont{Felser, Finckh,
  Kleinke, and Tremel}}]{felser98}
\bibinfo{author}{\bibfnamefont{C.}~\bibnamefont{Felser}},
  \bibinfo{author}{\bibfnamefont{E.}~\bibnamefont{Finckh}},
  \bibinfo{author}{\bibfnamefont{H.}~\bibnamefont{Kleinke}}, \bibnamefont{and}
  \bibinfo{author}{\bibfnamefont{W.}~\bibnamefont{Tremel}},
  \bibinfo{journal}{J. Mater. Chem.} \textbf{\bibinfo{volume}{8}},
  \bibinfo{pages}{1787} (\bibinfo{year}{1998}).

\bibitem[{\citenamefont{St\"owe and Wagner}(1998)}]{stowe98}
\bibinfo{author}{\bibfnamefont{K.}~\bibnamefont{St\"owe}} \bibnamefont{and}
  \bibinfo{author}{\bibfnamefont{F.}~\bibnamefont{Wagner}},
  \bibinfo{journal}{J. Solid State Chem.} \textbf{\bibinfo{volume}{138}},
  \bibinfo{pages}{160} (\bibinfo{year}{1998}).

\bibitem[{\citenamefont{Yokoya et~al.}(2005)\citenamefont{Yokoya, Kiss,
  Chainani, Shin, , and Yamaya}}]{yokoya05}
\bibinfo{author}{\bibfnamefont{T.}~\bibnamefont{Yokoya}},
  \bibinfo{author}{\bibfnamefont{T.}~\bibnamefont{Kiss}},
  \bibinfo{author}{\bibfnamefont{A.}~\bibnamefont{Chainani}},
  \bibinfo{author}{\bibfnamefont{S.}~\bibnamefont{Shin}}, , \bibnamefont{and}
  \bibinfo{author}{\bibfnamefont{K.}~\bibnamefont{Yamaya}},
  \bibinfo{journal}{Phys. Rev. B} \textbf{\bibinfo{volume}{71}},
  \bibinfo{pages}{140504(R)} (\bibinfo{year}{2005}).

\bibitem[{\citenamefont{Chung et~al.}(1993)\citenamefont{Chung, Wang, and
  Brill}}]{chung93}
\bibinfo{author}{\bibfnamefont{M.}~\bibnamefont{Chung}},
  \bibinfo{author}{\bibfnamefont{Y.}~\bibnamefont{Wang}}, \bibnamefont{and}
  \bibinfo{author}{\bibfnamefont{J.~W.} \bibnamefont{Brill}},
  \bibinfo{journal}{Synthetic Metals} \textbf{\bibinfo{volume}{55-57}},
  \bibinfo{pages}{2755} (\bibinfo{year}{1993}).

\bibitem[{\citenamefont{Zwick et~al.}(1980)\citenamefont{Zwick, Renucci, and
  Kjekshus}}]{zwick80}
\bibinfo{author}{\bibfnamefont{A.}~\bibnamefont{Zwick}},
  \bibinfo{author}{\bibfnamefont{M.}~\bibnamefont{Renucci}}, \bibnamefont{and}
  \bibinfo{author}{\bibfnamefont{A.}~\bibnamefont{Kjekshus}},
  \bibinfo{journal}{J. Phys. C, Solid St. Phys.} \textbf{\bibinfo{volume}{13}},
  \bibinfo{pages}{5603} (\bibinfo{year}{1980}).

\bibitem[{\citenamefont{Baker et~al.}(1978)\citenamefont{Baker, Nickel, and
  Meiron}}]{baker78}
\bibinfo{author}{\bibfnamefont{G.~A.} \bibnamefont{Baker}},
  \bibinfo{author}{\bibfnamefont{B.~G.} \bibnamefont{Nickel}},
  \bibnamefont{and} \bibinfo{author}{\bibfnamefont{D.~I.}
  \bibnamefont{Meiron}}, \bibinfo{journal}{Phys. Rev. B}
  \textbf{\bibinfo{volume}{17}}, \bibinfo{pages}{1365} (\bibinfo{year}{1978}).

\end{thebibliography}

\end{document}